\documentclass[12pt]{article}
\usepackage{graphicx}
\usepackage{amssymb}
\usepackage{amsmath}
\usepackage{xcolor}
\usepackage{soul}
\usepackage{cite}
\usepackage{mathrsfs}
\usepackage{tikz}
\usepackage{cancel}
\usepackage{mathtools}

\usepackage{array}

\makeatletter
\makeatletter

\DeclareMathOperator{\sgn}{sgn}

\newcommand{\rc}{\textcolor{red}}

\begin{document}

\begin{titlepage}
\vfill
\begin{flushright}
\end{flushright}

\vfill
\begin{center}
\baselineskip=16pt
{\Large\bf 
On the central singularity of the BTZ geometries}
\vskip 3mm
{\bf Mat\'{\i}as Brice\~no${}^{a}$, Cristi{\'a}n Mart\'{\i}nez${}^{b,c}$ and Jorge Zanelli${}^{b,c}$} \\
\vskip 6mm
{${}^a$ Instituto de F\'{\i}sica, Pontificia Universidad Cat{\'o}lica de Chile, Casilla 306, Santiago, Chile\\
${}^b$ Centro de Estudios Cient\'{\i}ficos (CECs), Av. Arturo Prat 514, Valdivia, Chile \\
${}^c$ Facultad de Ingenier\'{\i}a, Arquitectura y Dise\~no, Universidad San Sebasti\'an, \\sede Valdivia, General Lagos 1163, Valdivia 5110693, Chile
\vskip 0.2cm
\texttt{\footnotesize{matias.briceno.97@gmail.com, cristian.martinez@uss.cl, jorge.zanelli@uss.cl}}
} 
\vspace{6pt}\\
\today
\end{center}
\vskip 0.2in
\par
\begin{center}
{\bf Abstract}
\end{center}
\begin{quote}
The nature of the central singularity  of the BTZ geometries --stationary vacuum solutions of 2+1 gravity with negative cosmological constant $\Lambda=-\ell^{-2}$ and $SO(2)\times \mathbb{R}$ isometry-- is discussed. The essential tool for this analysis is the holonomy operator on a closed path (i.e., Wilson loop) around the central singularity. The study considers the holonomies for the Lorentz and AdS$_3$ connections. The analysis is carried out for all values of the mass $M$ and angular momentum $J$, namely, for black holes ($M \ell \ge |J|$) and naked singularities ($M \ell < |J|$). In general, both Lorentz and AdS$_3$ holonomies are nontrivial in the zero-radius limit revealing the presence of delta-like singularity at the origin in the curvature and torsion two-forms. However, in the cases $M\pm J/\ell=-n_{\pm}^2$, with $n_{\pm} \in \mathbb{N}$, recently identified in \cite{GMYZ} as BPS configurations, the AdS$_3$ holonomy reduces to the identity. Nevertheless, except for the AdS$_{3}$ spacetime ($M=-1$, $J=0$), all BTZ geometries have a central singularity which is not revealed by local operations. 
\vfill
\vskip 2.mm
\end{quote}
\end{titlepage}
\tableofcontents
\newpage
\section{Introduction}  
The BTZ geometries are exact stationary solutions of Einstein's equations in 2+1 dimensions with negative cosmological constant $\Lambda = -1/\ell^2$. They are described by the metric
\begin{equation} \label{BTZ}
ds^2= -\left(\frac{r^2}{\ell^2}-M\right)dt^2 -J dt d\theta + \left(\frac{r^2}{\ell^2}-M+\frac{J^2}{4r^2}\right)^{-1}dr^2+r^2 d\theta^2\,,
\end{equation}
where $-\infty < t< +\infty$, $0< r< \infty$, $0\leq \theta \leq 2\pi \cong 0$. This family is characterized by two parameters identified as the mass $M$ and angular momentum $J$, corresponding to the conserved charges associated to time-translation and rotational isometries, respectively.

Depending on the values of $M$ and $J$, different geometries arise from \eqref{BTZ}. The region $M \ell \geq |J|$ in parameter space $(M,J)$ corresponds to the  BTZ black hole \cite{BTZ, BHTZ}, with event ($+$) and Cauchy ($-$) horizons located at $r_\pm = |\sqrt{M+J/\ell}\pm \sqrt{M-J/\ell}|\ell/2$. For $M\ell < |J|$, the geometries \eqref{BTZ} have no horizons and are therefore naked singularities: conical in the case $M \ell \leq -|J|$ \cite{MiZ}, and \emph{overspinning geometries} \cite{BMZ} if $|M| \ell < |J|$. For $M\ell=|J|\neq 0$ and $M\ell=-|J|\neq 0$ these are extremal black holes and extremal conical singularities, respectively. Finally, for $M\ell=|J|=0$ these geometries reduce to the massless and spinless BTZ spacetime. The naked singularities appearing in the cases $M\pm J/\ell=-n_{\pm}^2$, with $n_{\pm} \in \mathbb{N}$, have been recently identified as BPS configurations \cite{GMYZ}. In addition, it has been shown \cite{BMZ} that closed timelike curves do not occur in these geometries for any values of $M$ or $J$ in the region $r^2>0$. The surfaces at fixed time in the geometries described by \eqref{BTZ}, have a singularity at the point $r=0$. The purpose of this work is to analyze the nature of this singularity.
\subsection{Identifications} 
The BTZ geometries described by the line element \eqref{BTZ} are quotients AdS$_3/\boldsymbol{K}$, where AdS$_3$ is the maximally symmetric, constant negative curvature manifold in 2+1 dimensions and $\boldsymbol{K}$ is a globally defined \textit{spacelike} Killing vector in the quotient space. The isometry group of AdS$_3$, $SO(2,2)$, has rank two --its Cartan subalgebra has two commuting generators. Therefore, identification Killing vectors given by linear combinations of two commuting generators produce geometries parametrized by two independent coefficients. The resulting BTZ geometries are described by a two-parameter family of metrics labeled by the real constants of integration $M$ and $J$ \cite{BHTZ}, which are functions of the coefficients of the linear combination of generators in the Killing vector. The identifications are summarized in Table \ref{table:tablevectors}. As this table shows, a black hole is obtained from two commuting boosts, conical singularities are generated by two commuting rotations, and combinations of a boosts and rotations result in overspinning spacetimes.

There are two inequivalent ways to define conical defects. One is the intuitive notion associated to the excision of a circular sector from a disc to form a cone by identifying the radii that define the sector. This produces a cone of any angular deficit $\Delta$, with $0\leq \Delta <2\pi$. The second approach is to define the conical manifold as a quotient through identification by a Killing vector field. In this case, all points in the manifold that can be connected by the action of a vector are identified. The problem arises when the Killing vector identifies points along a compact coordinate, say, $S^1$. If this notion is applied to construct a two-dimensional cone, the radius for $\theta=0$ is identified with the radius with $\theta=\theta_0$. But the action of the Killing field also identifies the radius with $\theta=\theta_0$ and that with $\theta=2\theta_0$, and $\theta=3\theta_0$, and so forth. If there is no pair of integers $n,m$ for which $n\theta_0=2m\pi$ this process does not terminate, identifying an infinite set of points densely distributed in the circle, which clearly does not yield a two-dimensional cone. But there is an additional problem. Consider the identification by the Killing vector $\boldsymbol{K} = \lambda\mathbf{J}$, where $\mathbf{J}$ is a generator of rotations and $\lambda=p/q$ is a rational number. Then, the conical manifold generated by this vector is indistinguishable from one generated by $\boldsymbol{K'} = \lambda'\mathbf{J}$ where $\lambda'=p'/q$, with $(p,q)=1= (p',q)$ (coprimes). For example, if $\lambda=1/7$ and $\lambda'=2/7$, these two vectors give rise to a cone with the same deficit angle of $\Delta=2\pi(1-1/7)=12\pi/7$ (the same manifold would also result if $\lambda'=3/7$, 4/7, 5/7 or 6/7). Therefore, in the construction of a cone by identification, it is sufficient to consider only $\lambda=1/n$ with $n \in \mathbb{N}$.

For angular excesses there are also two distinct notions to consider. The first corresponds to the insertion of an arbitrary angular sector, generating an Elizabethian geometry of arbitrary angular size $\theta_0>2\pi$. In this way, the angular periodicity is changed from $2\pi$ to $2\pi \lambda$ with $\lambda>1$. In the second approach, identifying points in a circle by a vector with a rational $\lambda=p/q>1$ always leads to a conical manifold with angular deficit $\Delta=2\pi(1-1/q)$. The analysis in this work applies independently of which notion is used. Since the Killing vector determines the mass and angular momentum of the resulting geometry, and these integration constants can in principle take any real value, the identification parameters are taken to be the most general ones.

In the case of the black hole spacetime ($M\ell>J$) this problem does not arise because the associated Killing vector corresponds to a boost, whose integral curve is noncompact on the pseudosphere, unlike the conical case. For the overspinning spacetime, the problem does not arise either for the same reason.

\begin{table}[ht!] \label{table:tablevectors}
\centering 
\scalebox{0.9}{
\begin{tabular}{c|c|c}
\hline
 \bf{Roots of $g^{rr}=0$} & \bf{Killing vector $\boldsymbol{K}$ } & \bf{Geometry} \\[1.5mm] \hline \hline 
$\lambda_+\in \mathbb{R}^+\,, \lambda_{-} \in \mathbb{R}$ & $\lambda_+ \mathbf{J}_{0 1} + \lambda_- \mathbf{J}_{23}$ & Generic BH,  $M\ell > |J|$ \\[1.5mm]
\hline 
$\lambda_{+}=|\lambda_{-}| \in \mathbb{R}^+$ & $\pm\lambda_+(\mathbf{J}_{01} + \mathbf{J}_{23})\pm\frac{1}{2}(\mathbf{J}_{02}+\mathbf{J}_{03}+\mathbf{J}_{12}+\mathbf{J}_{13})$& Extremal BH, $\pm M\ell = J$ \\[1.5mm]
\hline 
 $i\lambda_+ \in i\mathbb{R}^+ \, , i\lambda_{-} \in i \mathbb{R}$ & $\lambda_+ \mathbf{J}_{21}+\lambda_- \mathbf{J}_{30}$ & Generic CS, $M\ell < -|J|$ \\[1.5mm] \hline 
$i\lambda_+= i|\lambda_-| \in i \mathbb{R}^+$ & $\pm \lambda_+(\mathbf{J}_{03}-\mathbf{J}_{12})\mp\frac{1}{2} (\mathbf{J}_{01}+\mathbf{J}_{03}+\mathbf{J}_{12}-\mathbf{J}_{23})$ & Extremal CS, $\pm(-M)\ell = J$  \\[1.5mm] \hline 
$\lambda_{\pm}=0$ & $-\mathbf{J}_{12}+\mathbf{J}_{13}$  & $M\ell= 0 = J$ \\[1.5mm] \hline
$\lambda_{\pm}=a \pm ib$ & $b(\mathbf{J}_{03}+\mathbf{J}_{21})-a(\mathbf{J}_{01}+\mathbf{J}_{32})$ & Overspinning spacetime, \\ & $a=\text{sign}(J)\frac{\sqrt{|J|/\ell+ M}}{2},\; b= \text{sign}(J)\frac{\sqrt{|J|/\ell- M}}{2}$ & $-|J|< M\ell < |J|$ \; \\ [1.5mm] \hline
\end{tabular} }
\caption{Roots of $g^{rr}=\lambda^2 - M + J^2/(4 \lambda^2 \ell^2) =0$ (with $\lambda=r/\ell$) and identification Killing vectors $\boldsymbol{K}$ in terms of $SO(2,2)$ generators for each of the six BTZ geometries, where $\lambda_{\pm}= ( \sqrt{|M|+J/\ell} \pm \sqrt{|M|-J/\ell} )/2$ in the cases $ |M|\ell \geq |J|$, otherwise,  $\lambda_{\pm}=\text{sign}(J)( \sqrt{|J|/\ell+M} \pm i \sqrt{|J|/\ell-M} )/2$  for the  overspinning spacetimes. As discussed above, for the conical singularities, $M\ell \leq -|J|$, the identification parameters can be restricted to $\lambda_{\pm}=1/n_{\pm}$ with $n_{\pm}\in \mathbb{N}$. We assume the $\mathbb{R}^{(2,2)}$ metric $\eta_{AB}=\text{diag}(-1,+1,+1,-1)$. }
\end{table}

\subsection{Nature of the singularity} 
In most of the four- (and higher-) dimensional black holes, the Kretschmann invariant diverges as $r \to 0$ marking the presence of a curvature singularity at $r=0$. The metric (\ref{BTZ}) also has a singularity at the origin but its nature is puzzling because for $r\neq 0$ the geometry has constant negative curvature, which remains finite in the limit $r \to 0$. The central singularity in the BTZ geometries is therefore different from other black hole curvature singularities \cite{CLM,PRR,PL} and has not been fully understood. As discussed in \cite{BMZ}, the singularity in the BTZ spacetimes with $(M,J)\neq(-1, 0)$ is quasi-regular, characterized by having a finite Riemann curvature tensor in the neighborhood of the singularity. An early example of this class of singularities was presented in \cite{BrunoEllis}. These spacetimes do not have a well-defined tangent space at the singularity, which implies that they do not admit an extension beyond the singularity \cite{EllisSchmidt,Clarke}. This analysis can be replicated for all BTZ spacetimes. 

From a geometrical point of view, the question about the nature of the singularity still remains: What happens in the vicinity of $r=0$? How is the affine structure --the notion of parallel transport-- affected by the presence of the singularity? Does the curvature tensor present a Dirac delta with support at $r=0$ acting as a source? It was argued in \cite{BHTZ} that the non-vanishing holonomies of spacetime can be related to the nontrivial topology rather than to the existence of a source. Here we explore the complementary path that relates a nontrivial holonomy to the presence of a localized source. These two points of view are in fact equivalent, as emphasized by Misner and Wheeler \cite{Wheeler}. For an external observer a nontrivial topological structure can be indistinguishable from a localized source. The remaining of this article addresses these questions. The key for the analysis is the study of holonomies around $r=0$ in order to gauge the effect of parallel transporting a tangent space vector on a small closed loop around the singularity.

\section{Holonomies}  
The existence and nature of the singularity at the origin can be studied analyzing the holonomies that characterize the parallel transport of a vector in a closed loop around the singularity. These holonomies can be associated to the Lorentz connection or --considering the Chern-Simons formulation for three-dimensional gravity--, to the AdS$_3$ connection, as was discussed for the BTZ black hole in different contexts \cite{CLM,Vaz-Witten, Fernando:1998eg, Rooman:2000zi, Mansson:2000sj, Cho}. Here, these holonomies are studied for all geometries described by \eqref{BTZ} with $(M,J)\in \mathbb{R}^2$. 

\subsection{Parallel transport: Wilson loop for a Lorentz connection} \label{2.1} 
In order to determine the effect of parallel transporting a vector $u^\mu$ along a curve $x^{\mu}(\lambda)$ around the origin of a BTZ geometry, consider the parallel transport equation
\begin{equation}
\frac{d u^\mu}{d \lambda}+u^\alpha \Gamma^{\mu}_{\alpha \beta}\frac{d x^\beta}{d \lambda} =0.
\end{equation}
Note that $\Gamma^\mu_{\alpha \beta}$ is not necessarily symmetric in its lower indices, which allows for the presence of torsion in the geometry. Expressing $u^\mu$ in an orthonormal basis, $u^\mu=u^a e_a^{\mu}$, this equation can be written as
\begin{equation} \label{transport00}
0=\frac{d u^\mu}{d \lambda}+u^\alpha \Gamma^{\mu}_{\alpha \beta}\frac{d x^\beta}{d \lambda}= e_a^{\mu} \left(\frac{d u^a}{d \lambda}+\omega^{a}_{ \; \; b \,\alpha}  u^b \frac{d x^\alpha}{d \lambda}\right),
\end{equation}
where $\omega^{a}{}_{b\,\alpha}$ is the Lorentz (``spin") connection
\begin{equation} \label{spinconnection}
\omega^a{}_{b \mu} = e^a{}_\lambda [\partial_\mu e^\lambda{}_b +\Gamma^\lambda_{\mu \nu} e^\nu{}_b]. 
\end{equation}
Assuming invertibility of $e^\mu{}_a$, we have the equivalent version in the orthonormal frame,\footnote{Since the affine connection allows for torsion, $\omega^a{}_b$ is not necessarily torsion-free.}
\begin{equation} \label{transport01}
\frac{d u^a}{d \lambda}+\omega^{a}_{ \; \; b \,\alpha} u^b \frac{d x^\alpha}{d \lambda}=0,
\end{equation}
which expresses the condition for parallel transport of $u^a$ along a path, where the notion of parallelism is defined by the Lorentz connection $\omega$. Integrating this equation along a path $\mathcal{C}$, connecting points $x_1$ and $x_2$, relates the vector components $u^a$ at both ends,
\begin{eqnarray}\nonumber
u^a(x_2) &=& \lim_{N\to \infty}[\delta^a_{b_N} - \omega^a{}_{b_N}(x_N)][\delta^{b_N}_{b_{N-1}}- \omega^{b_N}{}_{b_{N-1}}(x_{N-1})]\cdots[\delta^{b_1}_b - \omega^{b_1}{}_b(x_1)] u^b(x_1) \\
& \equiv &\left[\mathbb{T}_L(x_2;x_1) \right]^a{}_b u^b(x_1)\;.
 \label{Gral.holonomy}
\end{eqnarray}
Here $\mathbb{T}_L$, the operator of parallel transport between $x_1$ and $x_2$, is given by the Lorentzian Wilson line,
\begin{equation} \label{T-generic}
\mathbb{T}_L(x_2;x_1) = P_{\mathcal{C}}    \exp\left[-\int_{x_1} ^{x_2} \omega\right],
\end{equation}
where $P_{\mathcal{C}}$ denotes the path-ordered product of the infinitesimal decomposition of the exponential along the path $\mathcal{C}$. 

If the path $\mathcal{C}$ happens to be the integral curve of a Killing vector, the connection can be chosen to be constant along the path. This makes the ordering prescription unnecessary, and the integral in \eqref{T-generic} reduces to the product of the connection times the length of the path. The stationary geometries described by \eqref{BTZ} have two global Killing vectors, $\partial_t$ and $\partial_\theta$. Since we are interested in closed paths around $r=0$, the $P_{\mathcal{C}}$-prescription is irrelevant on circular or helical paths with $dr=0$, and the parallel transport operator takes the form
\begin{equation} \label{TL}
\mathbb{T}_L(\lambda) = \exp[-\lambda \omega_{\lambda}],
\end{equation}
where $[\omega_\lambda]^a{}_b = \omega^a{}_{b \mu} (dx^\mu/d \lambda) $ and $\lambda$ is the parameter along the curve. Thus, the parallel transport of a vector $u^a$ on a closed loop around $r=0$ can be determined either by integrating the first order equation of parallel transport \eqref{transport01}, or by computing the holonomy \eqref{TL}. This second approach relies exclusively on the existence of the Killing vector $\boldsymbol{K}$ and can be applied to geometries like \eqref{BTZ} which are expressible as AdS$_3/\boldsymbol{K}$. This method has been used to classify gravitational instantons by studying the action of one-parameter groups of isometries in the vicinity of a curvature singularity \cite{Gibbons-Hawking}. The idea is to construct the holonomy around a fixed point of the Killing vector from its covariant derivative, to examine the parallel transport on a small loop around the singularity, giving the same result (\ref{TL}).
\subsubsection{Positive mass BTZ spacetimes} 
Consider the positive mass BTZ spacetimes --which include the 2+1 black hole geometry--  defined by the {\it dreibein}
\begin{equation} \label{e-M>0}
e^0= \frac{\sqrt{B}}{\sqrt{-A}} \;d\theta, \qquad e^1= \frac{r dr}{\sqrt{B}}, \qquad e^2=\sqrt{-A}dt-\frac{J}{2\sqrt{-A}}d\theta\,, 
\end{equation}
where $A$ and $B$ depend only on the square of the radial coordinate $r$,  
\begin{equation} \label{onshell}
 A=-M+\frac{r^2}{\ell^2} \qquad \text{and} \qquad B = \frac{r^4}{\ell^2}-M r^2 +\frac{J^2}{4}\,.
\end{equation}
The local frame \eqref{e-M>0} is well defined in an open patch around the origin provided $M>0$ and $J\neq 0$: $A\simeq-M$, $B\simeq J^2/4$. (The corresponding local frame for other values of $M$ and $J$ requires a different dreibein). Assuming a vanishing torsion for $r \neq 0$, the Lorentz connection $\omega$ can be found either by solving the algebraic equation $T^a=d e^a + \omega^a{}_b \;e^b =0$, or by substituting the dreibein \eqref{e-M>0} in \eqref{spinconnection}, leading to
\begin{align} \label{w-M>0}
\omega^{01}=\frac{J}{2\ell^2 \sqrt{-A}}dt -\sqrt{-A}d\theta, \quad \omega^{02}= \frac{-J r}{2\ell^2 A \sqrt{B}}dr, \quad \omega^{12} = \frac{\sqrt{B}}{\ell^2 \sqrt{-A}}dt.
\end{align}

Using the Lorentz connection \eqref{w-M>0} in the limit $r\rightarrow 0$, the exponent in \eqref{TL} is given by
\begin{equation}
\boldsymbol{ \omega_{M>0}}:= [\omega_{ \theta}]^{a}_{ \; \; b }=\left(
\begin{array}{ccc}
 0 & -\sqrt{M} & 0 \\
 -\sqrt{M} & 0 & 0 \\
 0 & 0 & 0 \\
\end{array}
\right),
\end{equation}
and therefore, the parallel transport operator \eqref{TL} for $r \to 0$ is 
\begin{equation} \label{map+}
\mathbb{T}^{M>0}_L(\theta)=\exp(-\theta \, \boldsymbol{ \omega_{M>0}})=\left(
\begin{array}{ccc}
 \cosh \left(\sqrt{M}\theta  \right) & \sinh \left( \sqrt{M} \theta \right) & 0 \\
 \sinh \left(\sqrt{M}\theta  \right) & \cosh \left(\sqrt{M}\theta  \right) & 0 \\
 0 & 0 & 1 \\
\end{array}
\right).
\end{equation}
Note that in the case $J=0$ the dreibein (\ref{e-M>0}) is not well defined for $r \to 0$, but choosing the appropriate frame 
\begin{align}
    e^0=-\frac{dr}{\sqrt{-A}}, \quad e^1= rd\theta, \quad e^2=\sqrt{-A}dt,
\end{align}
the holonomy close to $r=0$ still yields the same result (\ref{map+}). Note that the operator (\ref{map+}) is not a periodic function of $\theta$, nor is it the identity for $\theta=2\pi$. This map represents a finite boost in the $(0-1)$ plane that relates a vector after parallel transport around $r=0$ back to the starting point to the original vector. This happens no matter how small the loop is, which indicates the existence of a nontrivial curvature flux concentrated at $r=0$. 

\subsubsection{Negative mass BTZ spacetimes}

The local frame in the case of negative mass $M$ can be chosen as 
\begin{equation} \label{e-M<0}
e^0=\sqrt{A}dt +\frac{J}{2\sqrt{A}}d\theta,\qquad e^1= \frac{r dr}{\sqrt{B}},\qquad e^2= \frac{\sqrt{B}}{\sqrt{A}} \;d\theta, 
\end{equation}
with $A$ and $B$ given in \eqref{onshell}, which is regular in the vicinity of the origin provided $M<0$. The corresponding torsion-free spin connection is now found to be
\begin{align} \label{w-M<0}
\omega^{01}=\frac{\sqrt{B}}{\ell^2\sqrt{A}}dt,\quad \omega^{02} = \frac{J r }{2 A \ell^2 \sqrt{B}}dr,\quad \omega^{12}= -\frac{J}{2\ell^2 \sqrt{A}}dt -\sqrt{A}d\theta.
\end{align}

From \eqref{w-M<0}, in the limit $r \to 0$, $\omega_{ \theta}$ is
\begin{equation}
\boldsymbol{ \omega_{M<0}}\equiv [\omega_{ \theta}]^{a}_{ \; \; b } =\left(
\begin{array}{ccc}
 0 & 0 & 0 \\
 0 & 0 & -\sqrt{-M} \\
 0 & \sqrt{-M} & 0 \\
\end{array}
\right).
\end{equation}
Thus, the exponential map in the limit $r \to 0$ is given by
\begin{align} \label{map-}
\mathbb{T}^{M<0}_L(\theta):=\exp(-\theta \, \boldsymbol{ \omega_{M<0}})=\left(
\begin{array}{ccc}
 1 & 0 & 0 \\
 0 & \cos \left(\sqrt{-M}\,\theta  \right) & \sin \left(\sqrt{-M}\,\theta  \right) \\
 0 & -\sin \left(\sqrt{-M}\,\theta \right) & \cos \left(\sqrt{-M}\,\theta  \right) \\
\end{array}
\right).
\end{align}

This is a rotation matrix by an angle $\sqrt{-M} \theta$ in the spatial plane. This means that after parallel transport along a closed infinitesimal loop around the origin, a vector comes back rotated by an angle $2\pi \sqrt{-M}$ with respect to the original vector in the plane tangent to the spatial slice. The fact that for a generic $M$ the matrix $\mathbb{T}^{M<0}_L(2\pi)$ is not the identity is due to the presence of a nontrivial defect concentrated at $r=0$. 

For $0>M>-1$ and $J=0$, BTZ spacetimes contain a deficit angle \cite{MiZ} given by
\begin{equation} \label{defect}
\Delta= 2\pi(1-\sqrt{-M})\,.
\end{equation}
Note that the angular defect is related to the identification Killing vector $\boldsymbol{K} = \lambda_+\mathbf{J}_{21}$, where $\lambda_+=[1-\Delta/(2\pi)]=\sqrt{-M}$. As mentioned before, a peculiarity of the angular identification in a compact space, is that $\lambda_+=p/q$ and $\lambda'_+=p'/q$ --with $(p,q)=1= (p',q)$ ($p$ and $q$ coprimes, that is, without common factors)--, give rise to the same manifold. This 
would mean that different values of the mass, say $M=-1/q^2$ and $M'=-(p/q)^2$, would correspond to the same conical manifold. In what follows this peculiarity will be ignored as it does not affect the conclusions of this work.

For $M=-1, J=0$, the geometry \eqref{BTZ} is anti-de Sitter space, $\mathbb{T}^{M=-1,J=0}_L(2 \pi)=\mathbf{1}$ and there is no deficit angle (no conical defect). If $M<-1$ (negative $\Delta$), the geometry has an angular excess. In the special cases $M=-n^2$, $n \geq 2\in \mathbb{N}$, the transported vector is rotated by $2\pi n$ and the holonomy (\ref{map-}) equals the identity, even though there is still curvature concentrated at $r=0$. These cases correspond to $n$-fold coverings of AdS$_3$ and are also BPS states \cite{GMYZ}.  

\subsubsection{$M=0$ and $J \neq 0$ spacetimes} 

Next, the parallel transport operator is determined for the massless overspinning  ($M=0$, $J \neq 0$) BTZ geometries, described by the stationary line element
\begin{equation} \label{BTZmetric}
ds^2= -\frac{r^2}{\ell^2}dt^2 -Jdt d\theta +\frac{r^2 \ell^2}{r^4+\frac{J^2 \ell^2}{4}}dr^2+r^2 d\theta^2,
\end{equation}
where the coordinates  ranges are $-\infty<t<\infty$, $0<r<\infty$, and $0\leq \theta\leq 2\pi$. The study requires a regular frame in the limit $r\to 0$. The 
 dreiben and spin connection can be chosen as
\begin{align}\label{frameM0}
e^0&=\frac{\sqrt{J^2+r^2}}{\ell}dt +\frac{J\ell \left( \sqrt{J^2+r^2}- \sqrt{J^2+4 r^4/\ell^2}\right)}{2 r^2}d\theta,\quad e^1= \frac{2 \ell r dr}{\sqrt{J^2 \ell^2+4 r^4}}, \nonumber\\ e^2&= -\frac{J}{\ell} dt+\frac{\sqrt{J^2+r^2} \sqrt{J^2 \ell^2+4 r^4}-J^2 \ell}{2 r^2} d\theta, 
\end{align}
\begin{align}\label{frame-sc-M0w01}
\omega^{01}&=\frac{\sqrt{J^2+r^2} \sqrt{J^2 \ell^2+4 r^4}-J^2 \ell}{2 \ell^2 r^2}dt -\frac{J}{\ell} d\theta,\\ \omega^{02}&=\frac{J}{r} \left(\frac{1}{\sqrt{J^2 +4 r^4/\ell^2}}-\frac{1}{\sqrt{J^2+r^2}}\right)dr,\\ \omega^{12}&=\frac{J \sqrt{J^2 +4 r^4/\ell^2}-J \sqrt{J^2+r^2}}{2 \ell r^2}dt -\frac{\sqrt{J^2+r^2}}{\ell}d\theta. \label{frame-sc-M0w12}
\end{align}

From (\ref{frame-sc-M0w01}) - (\ref{frame-sc-M0w12}), in the limit $r\to 0$ the Lorentz connection is
\begin{equation}
\boldsymbol{ \omega_{M=0}}\equiv [\omega_{ \theta}]^{a}_{ \; \; b } =\left(
\begin{array}{ccc}
0 & -\frac{J}{\ell} & 0 \\
-\frac{J}{\ell} & 0 & -\frac{|J|}{\ell} \\
0 & \frac{|J|}{\ell} & 0 \\
\end{array}
\right),
\end{equation}
that corresponds to a rotation in the $(1-2)$ plane together with a boost in the $(0-1)$ plane. This is a nilpotent matrix of third degree ($\boldsymbol{ \omega_{M=0}}^3=0$) and therefore
\begin{align}
\mathbb{T}^{M=0}_L(\theta):=\exp(-\theta \, \boldsymbol{ \omega_{M=0}})=\left(
\begin{array}{ccc}
 \frac{\theta ^2 J^2}{2 \ell^2}+1 & \frac{\theta  J}{\ell} & \frac{\theta ^2 J |J|}{2 \ell^2} \\
 \frac{\theta  J}{\ell} & 1 & \frac{\theta  |J|}{\ell} \\
 -\frac{\theta ^2 J |J|}{2 \ell^2} & -\frac{\theta  |J|}{\ell} & 1-\frac{\theta ^2 J^2}{2 \ell^2} \\
\end{array}
\right). \label{holonomyM0Jfix}
\end{align}
This operator is neither periodic in $\theta$ nor is the identity for $\theta=2 \pi$. It gives a finite element of the $SO(2,1)$ group that reduces to the identity in the limit $J\to 0$.

\subsubsection{$M=0$ and $J=0$ spacetime} 

For the black hole vacuum spacetime, $M=0=J$, a new radial coordinate $u$ defined as $u=\log(r/\ell)$ provides a suitable metric  to analyze the region $r\to 0$, which is given by
\begin{align} \label{newMJ=0}
    ds^2=-e^{2u}dt^2 +\ell^2 du^2 + \ell^2 e^{2u}d\theta^2,
\end{align}
where $u \in (-\infty,\infty)$. In this coordinate system, the dreibein can be chosen as 
\begin{align} \label{eMJ=0}
    e^0=e^{u}dt, \quad e^1= \ell du, \quad e^2 =\ell e^{u}  d\theta 
\end{align}
and the spin connection is given by
\begin{align} \label{wMJ=0}
    \omega^0_{\, \, \,1}=\frac{e^{u}}{\ell}dt, \quad \omega^{0}_{\, \, \, 2}=0,  \quad \omega^1_{\, \, \, 2}=-e^{u}d\theta.
\end{align}
The metric, the dreibein and spin connection can be obtained by taking the limit $J\to0$ of equations \eqref{BTZmetric}-\eqref{frame-sc-M0w12} for fixed $r$. Hence, the parallel transport operator \eqref{TL} reads
\begin{align}\label{Tmassless}
\mathbb{T}^{M=0=J}_L(\theta)=
\left(
\begin{array}{ccc}
 1 & 0 & 0 \\
 0 & \cos \left(e^u \theta\right) & \sin \left(e^u \theta\right) \\
 0 & -\sin \left(e^u \theta\right) & \cos \left(e^u \theta\right) \\
\end{array}
\right).
\end{align}
Although this operator does not have a period of $2\pi$, it reduces to the identity in the limit $u\to -\infty$, (i.e., $r\to 0$). Note that \eqref{Tmassless} is not the limit $J\to 0$ of \eqref{holonomyM0Jfix}. In this case the holonomy is a rotation that reduces to the identity in the limit $r\to 0$, whereas \eqref{holonomyM0Jfix} is a boost plus a rotation that approaches the identity as $J\to 0$. This highlights the non-commutativity of the limits $J\to 0$ and $r\to 0$. 

It is important to point out, nevertheless, that the geometry with $M=0=J$ is not singularity-free, but describes a conical defect with $\Delta=2 \pi$. However, an alternative point of view could be adopted in which the space is free of singularity but the topology is that of a cylinder. 

\subsection{Wilson loop for the AdS$_3$ connection}  \label{2.2} 
Three-dimensional gravity in the presence of a negative cosmological constant can be formulated as a Chern-Simons theory \cite{Witten}, where the gauge field is an $\mathfrak{so}(2,2)$ connection,
\begin{equation} \label{AdSconnection}
\mathbf{A} = (1/2) \omega^{ab} \mathbf{J}_{ab} + (e^a/\ell) \mathbf{J}_a\,,
\end{equation}
with $a,b=\{0,1,2\}$. Following the conventions in \cite{MiZ}, the six generators of $SO(2,2)$, $\mathbf{J}_{AB}=-\mathbf{J}_{BA}$, with $A=\{a,3\}$, $B=\{b,3\}$ and $\mathbf{J}_{a3}\equiv \mathbf{J}_a$, satisfy the algebra
\begin{equation} \label{algebra}
\left[ \mathbf{J}_{AB},\mathbf{J}_{CD}\right] =\eta _{AD}\,\mathbf{J}_{BC}-\eta
_{BD}\,\mathbf{J}_{AC}-\eta _{AC}\,\mathbf{J}_{BD}+\eta _{BC}\,\mathbf{J}_{AD},
\end{equation}
with $\eta_{AB}=\text{diag}(-1,1,1,-1)$.

The curvature associated with this AdS$_3$ connection contains information about both the Lorentz curvature ($R^{ab}$) and the torsion ($T^a$) two-forms,
\begin{equation}\label{F}
\mathbf{F}  = (1/2)(R^{ab} + e^a e^b/\ell^2)\mathbf{J}_{ab} + (T^a/\ell) \mathbf{J}_a\,.
\end{equation}
In absence of sources or other dynamical fields, the Chern-Simons equations, $\mathbf{F}=0$, imply that the geometry describes a locally flat AdS$_3$ manifold.  Hence, these on-shell locally AdS$_3$ geometries satisfy
\begin{equation} \label{AdS3-flat}
  R^{ab} + e^a e^b/\ell^2=0\,, \qquad   T^a=0,
\end{equation}
like those described by \eqref{BTZ} for $r\neq 0$. The BTZ solutions that satisfy \eqref{AdS3-flat} for $r\neq 0$, allow for $\delta$-like sources if the origin is included in the geometries. These sources can be determined from Wilson loops around the origin in the zero-radius limit of a circular loop. In this limit, the Wilson loop does not reduce to the identity {as shown below}, which indicates the presence of nonvanishing AdS$_3$ curvature at the origin.

In what follows, we consider Wilson loops for this $\mathfrak{so}(2,2)$ connection for the geometries described by \eqref{BTZ},
\begin{equation} \label{W-generic}
\mathcal{W}(\mathcal{C}) = P_{\mathcal{C}} \left[\exp{\left(-\oint \mathbf{A}\right)}\; \right],
\end{equation}
where $\mathcal{C}$ is the closed circular curve $\theta \in [0, 2\pi]$ with constant $t$ and $r$. An useful feature of this Wilson loop is the independence on the angular coordinate of the gauge connection. The equation for a Wilson line can be written as
\begin{align}
    d\mathcal{W}=-\mathbf{A} \, \mathcal{W}.
\end{align}
If the circle, which is the path in this case, is cut into $N$ small enough pieces, the above equation for each small piece reads 
\begin{align}
    \mathcal{W}(\theta_{i+1},\theta_{0})=(\mathbf{1}-\mathbf{A}_{\theta}(\theta_{i})d \theta_i)\mathcal{W}(\theta_{i},\theta_{0})
\end{align}
where $d\theta_i=\theta_{i+1}-\theta_i$ and $\mathcal{W}(\theta_i,\theta_0)$ is the Wilson line from $\theta_0$ to $\theta_i$ along the circular path. Assuming that $\mathcal{W}(\theta_0,\theta_0)=\mathbf{1}$, the Wilson line for a loop should be
\begin{align}
    \mathcal{W}(\theta_N,\theta_0)=(\mathbf{1}-\mathbf{A}_{\theta}(\theta_{N})d \theta_N)(\mathbf{1}-\mathbf{A}_{\theta}(\theta_{N-1})d \theta_{N-1})...(\mathbf{1}-\mathbf{A}_{\theta}(\theta_{1})d \theta_1).
\end{align}
Each factor in this product is approximately $e^{-\mathbf{A}_{\theta}(\theta_i)d\theta_i}$, since $d\theta_i$ is assumed to be sufficiently small for all $\theta_i$. Then, as $\partial_{\theta}$ is an isometry of the manifold, $\mathbf{A}_{\theta}$ can be chosen to be independent of $\theta$ and therefore 
\begin{align}
    \mathcal{W}(\theta_N,\theta_0)=e^{-\mathbf{A}_{\theta}d\theta_N} e^{-\mathbf{A}_{\theta}d\theta_{N-1}}...e^{-\mathbf{A}_{\theta}d\theta_1}=e^{-\mathbf{A}_{\theta}\sum_i d\theta_i} =e^{-2 \pi \mathbf{A}_{\theta}}\label{wilsontheta0}
\end{align}
where $\mathbf{A}_{\theta}$ is the $\theta$ component of the gauge connection. Thus, the Wilson loop \eqref{W-generic} takes the simple form
\begin{equation} \label{wilsontheta}
\mathcal{W}(\mathcal{C}) = \exp{\left[-2\pi \mathbf{A}_{\theta}\right]}.
\end{equation}
The next section is devoted to extract the AdS$_3$ curvature from \eqref{wilsontheta} in the cases where this Wilson loop is nontrivial ($\mathcal{W}(\mathcal{C})\neq \mathbf{1}$).

\section{Curvature at the center} \label{3}  
The Ambrose-Singer theorem \cite{Ambrose-Singer} relates the holonomy group with the curvature two-form enclosed by a path. In this sense, a nontrivial Wilson loop reveals the presence of a nonvanishing curvature. Computing an explicit expression for the AdS$_3$ curvature from the holonomy is in general a difficult problem due to the nonabelian character of the AdS$_3$ curvature. This calculation requires an suitable generalization of Stokes's theorem. This generalization was provided, for instance in \cite{ninobralic}, whose essential aspects are summarized below.

\subsection{Relation between the Wilson loop and curvature} \label{3.1} 

An explicit formula relating the Wilson operator and the curvature was already shown in \cite{ninobralic,Broda,Arefeva}. In this section the necessary ingredients to use this relation are introduced. Let $\mathcal{W}(\mathcal{C})$ be the Wilson line along the loop $\mathcal{C}$, which is the boundary of a two-dimensional surface $S$  parameterized with coordinates $\xi(\tau,s)$, and $\frac{\partial \xi^{\mu}}{\partial s}$, $\frac{\partial \xi^{\mu}}{\partial \tau}$, the tangent vectors of a given path along the $s$ and $\tau$ directions respectively. Moreover,  $\mathcal{X}(\gamma_{\tau}(s))$ denotes the Wilson line along the path $\gamma_{\tau}$ from $L=0$ to $L=s$ with $L$ being the parameter of the line. Then, the following expression holds,
\begin{align}\label{W-F}
    \mathcal{W}(\mathcal{C})=P_{\tau}\exp\left[-\int_{S} ds d\tau \frac{\partial \xi^{\mu}}{\partial s} \frac{\partial \xi^{\nu}}{\partial \tau}\mathcal{X}^{-1}(\gamma_{\tau}(s))\mathbf{F}_{\mu \nu}(\xi(\tau,s))\mathcal{X}(\gamma_{\tau}(s))\right].
\end{align}
The paths $\gamma$ are such that they start in a point on $\mathcal{C}$ and ends at a point $\xi(\tau,s)$. The details about the paths are described in  \cite{Arefeva,ninobralic}. Note that the components of the field strength $\mathbf{F}_{\mu \nu}$ are contracted with the tangent vectors over the surface and then integrated. This is equivalent to integrate the two-form curvature $\mathbf{F}$ over such surface. 

The formula (\ref{W-F}) is a bit involved since also contains the Wilson lines $\mathcal{X}(\gamma_{\tau}(s))$. However, the goal is to apply it in the limit in which the radius of the closed path $\mathcal{C}$ goes to zero. In this limit, the formula (\ref{W-F}) becomes simple and useful. 

If the two-form curvature is zero everywhere for BTZ spacetimes\footnote{This statement is equivalent to fulfill the Einstein field equations with a negative cosmological constant and to impose a vanishing torsion.}, all the Wilson loops are trivial as it follows from (\ref{W-F}). Then, the existence of a zero-radius non-trivial Wilson loop implies a topological defect emerging by the exclusion of $r=0$ in the manifold \cite{BHTZ}. As mentioned in the Introduction, there is an alternative point of view \cite{Wheeler},  in which such topological defect is equivalent to the  presence of a localized source in a background of trivial topology. This localized source is naturally chosen as a delta two-form, with support at $\Sigma$ which contains the origin, such that the two-form curvature has the following expression 
\begin{align}\label{Fform}
    \mathbf{F}=2\pi \mathbf{f} \delta(\Sigma), \quad \text{with} \quad \int_{S_0}\delta(\Sigma)=1,
\end{align}
where $S_0$  is a surface that intersects the line $r=0$ once and $\mathbf{f}$ is a linear combination of the generators of $SO(2,2)$. Additionally, in the limit in which the radius of $\mathcal{C}$ \rc{is} negligible compared to the AdS$_3$ length $\ell$, the Wilson lines between two distinct points reduce to the identity. This statement is expected in the case in which $\mathcal{C}$ shrinks around a regular point. However, the case of a loop enclosing a singular point requires a separate analysis if the Wilson line ends at the singularity. In what follows, it is shown  that those Wilson lines also reduce to the identity. 

\begin{figure}
    \centering
\begin{tikzpicture}
\draw (0,0) circle[radius=1.5];
\draw (0,0) circle[radius=2.5];
\draw (-1.76,1.76) .. controls (0,2.5) and (-2.5,0) .. (0,0);
\filldraw[black] (-1.2,0.88) circle (2pt) node[anchor=west]{$\Upsilon(s_r)$};
\filldraw[black] (-1.76,1.76) circle (2pt) node[anchor=east]{$\Upsilon(s_1)$};
\filldraw[black] (0,0) circle (2pt) node[anchor=west]{$\Upsilon(s_0)$};
\node at (2.8,0) {$\mathcal{C}$};
\node at (1.7,0.88) {$\mathcal{C}_r$};
\end{tikzpicture}
\caption{Path $\Upsilon(s)$ intersecting circles $\mathcal{C}$ and $\mathcal{C}_r$ at $\Upsilon(s_1)$ and $\Upsilon(s_r)$ respectively, with center at $\Upsilon(s_0)$.}
\label{fig:1}
\end{figure}
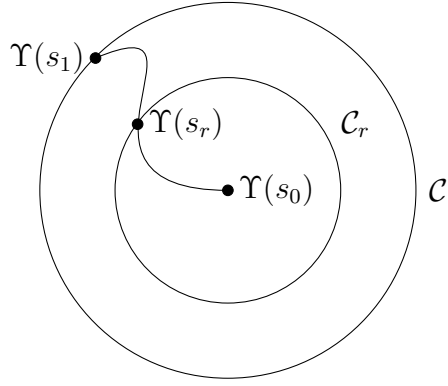

Consider a circular path $\mathcal{C}_{r}$ of radius $r$ as shown in Figure \ref{fig:1}. Let $\Upsilon(s)$ be a path parameterized by $s$ that ends at the singularity, where $\Upsilon(s_r)$ is the intersection between $\mathcal{C}_r$, and $\Upsilon(s_0)$.\footnote{As BTZ spacetimes are smooth or quasi-regular --in the sense that the singularity can be reached in a finite proper time--, the description above makes sense as a path that fulfills such conditions can always be found.} In the case of zero curvature, the gauge connection and the Wilson line to the singularity can always be written as
\begin{align}
    \mathbf{A}=-U^{-1}dU; \quad \mathcal{X}(\Upsilon(s_0))=U^{-1}(\Upsilon(s_0))U(\Upsilon(s_r)); \quad U\in SO(2,2).
\end{align}
In the limit of $r \to 0$, assuming that the path $\Upsilon(s)$ is smooth, then as $\Upsilon(s_{r})\to \Upsilon(s_0)$, the Wilson line $\mathcal{X}$ as shown above tends to the identity. This is the case as long as $U$ is defined in such limit, which happens for the BTZ spacetimes.

Since there is a single point in the integral (\ref{W-F}) that gives a nonzero contribution, the ordering prescription can be simplified and, in the limit $r\to 0$ of a shrinking circular path around the singularity, (\ref{W-F}) yields
\begin{align}\label{wilsonf}
    \mathcal{W}(\mathcal{C})=\exp{ \left[ -2 \pi \mathbf{f} \right]}.
\end{align}
Comparing  \eqref{wilsontheta} and \eqref{wilsonf} leads to $\exp(-2\pi \mathbf{f})=\exp(-2\pi \mathbf{A}_{\theta}{|}_{r=0})$, or equivalently,
\begin{align}\label{W=expA}
  \exp(-2\pi \mathbf{f})\exp(2\pi \mathbf{A}_{\theta}{|}_{r=0})=\mathbf{1},
\end{align}
where $\mathbf{A}_{\theta}{|}_{r=0}$ can be generically written as
\begin{align}\label{alpha_beta}
   \mathbf{A}_{\theta}{|}_{r=0}=\frac{1}{2}\alpha^{ab}\mathbf{J}_{ab}+\frac{\beta^{a}}{\ell}\mathbf{J}_{a}.
\end{align}
The equation (\ref{W=expA}) does not uniquely determine $\alpha^{ab}$ and $\beta^{a}$ in (\ref{alpha_beta}). In fact, this problem can be seen from the multiple roots of the identity, i.e. $\mathbf{1} = e^{2 \pi \mathbf{J_{03}}}=e^{2 \pi \mathbf{J_{12}}}=e^{2 \pi (n\mathbf{J_{03}}+m \mathbf{J}_{12})}$ with $n,m \in   \mathbb{Z}$. Consequently, this ambiguity leads to a fundamental indeterminacy (modulo an integer) of the coefficients $\alpha^{ab}$ and $\beta^a$ of the curvature and torsion. This problem will be further address in Section \ref{discusion}.

A sufficient condition to satisfy (\ref{W=expA}) is
\begin{align}\label{f=A}
    \mathbf{f}=\mathbf{A}_{\theta}{|}_{r=0}.
\end{align}
Then, from \eqref{F} and \eqref{f=A} the curvature and torsion of the manifold are obtained:
\begin{equation}\label{cur&tor}
    R^{ab}+\frac{1}{\ell^2}e^{a}e^{b}=2 \pi \alpha^{ab} \delta(\Sigma), \qquad
    T^{a}=2 \pi \beta^a \delta(\Sigma),
\end{equation}
where $\delta(\Sigma)$ stands for the 2-form delta distribution with support at $r=0$.

Additionally, one could construct an action principle for these equations using the first-order formalism, given by
\begin{align} \label{source}
    S=\int \epsilon_{abc}(R^{ab}e^{c}+\frac{1}{3 \ell^2}e^{a} e^{b}e^{c} - \alpha^{ab}e^{c}\delta(\Sigma)-\beta^{a}\omega^{bc}\delta(\Sigma)),
\end{align}
where the delta function is defined in (\ref{Fform}). It is interesting to observe that since $\delta(\Sigma)$ is in the $dr\wedge d\theta$ plane\footnote{In the coordinates of equation (\ref{BTZ})}, the term $\alpha^{ab}e^{c}\delta(\Sigma)+\beta^{a}\omega^{bc}\delta(\Sigma)$ is equal to $\alpha^{ab}e^{c}_{t}\;dt \wedge \delta(\Sigma)+\beta^{a}\omega^{bc}_{\;\;\;\;t}\; dt \wedge \delta(\Sigma)$ that is similar to the action considered in \cite{Vaz-Witten}.

\subsection{Curvature and torsion at $r=0$}\label{3.2}

In what follows the Lorentz curvature and torsion, which are the AdS$_3$ curvature components, are determined at the central singularity from the non-trivial Wilson loops as shown in the previous section.

For $M>0$,  the regular frame near $r=0$, given by Eqs. \eqref{e-M>0} and \eqref{w-M>0}, is used for building the AdS$_3$ connection \eqref{AdSconnection}. Then, from \eqref{wilsontheta} the Wilson loop at $r=0$ is obtained as 
\begin{align} \label{WilsonM>0}
    \mathcal{W}_{M>0}=\exp\left(-2\pi \left[ \frac{|J|}{2 \ell \sqrt{M}}\mathbf{J}_{03}-\frac{J}{2\ell \sqrt{M}}\mathbf{J}_{23}-\sqrt{M}\mathbf{J}_{01}\right] \right).
\end{align}  
This Wilson loop shows that there is an issue at $r=0$ since it does not reduce to the identity no matter how small the loop is. This means that there are AdS$_3$ curvature concentrated at $r=0$ and therefore this curvature has a $\delta$-like singularity supported at the origin. Using \eqref{cur&tor} and \eqref{WilsonM>0}, the nontrivial AdS$_3$ curvature components are identified as
\begin{align}
    R^{01}+\frac{e^{0}e^{1}}{\ell^2}=&-2\pi \sqrt{M}\delta(\Sigma)\\
    T^{0}=&\pi \frac{|J|}{ \sqrt{M}}\delta(\Sigma)\\
    T^{2}=& - \pi \frac{J}{ \sqrt{M}}\delta(\Sigma).
\end{align}
The above expressions can be considered as the gravitational field equations in the first-order formalism, where the right-hand side of them reveals the existence of a $\delta$-like source. 

The same procedure can be applied to the $M<0$ and $M=0$ cases. For $M<0$, the suitable frame is given by Eqs.  (\ref{e-M<0}) and (\ref{w-M<0}). Then, the Wilson loop at $r=0$ reads
\begin{align} \label{WM<0}
    \mathcal{W}_{M<0}=\exp\left(-2\pi \left[ \frac{J}{2 \ell \sqrt{-M}}\mathbf{J}_{03}+\frac{|J|}{2\ell \sqrt{-M}}\mathbf{J}_{23}-\sqrt{-M}\mathbf{J}_{12}\right]\right),
\end{align}
and the nonvanishing curvature components are
\begin{align}
    R^{12}+\frac{e^{1}e^{2}}{\ell^2}=&-2\pi \sqrt{-M} \delta(\Sigma) ,\label{negative_source} \\
    T^{0}=&\pi \frac{J}{ \sqrt{-M}}\delta(\Sigma),\\
    T^{2}=&  \pi \frac{|J|}{ \sqrt{-M}}\delta(\Sigma).
\end{align}
It is important to note that for $M=-1$ and $J=0$ --AdS$_3$ spacetime--, the Wilson loop \eqref{WM<0} reduces to the identity, as expected. However, the right hand side of \eqref{negative_source} does not vanish in that case. Indeed, for the static conical singularities ($M < 0$,
$J = 0$), the Wilson loop \eqref{WM<0} simply reads
\begin{align} \label{WMAdS0}
    \mathcal{W}_{M<0,J=0}=\exp\left(2\pi \sqrt{-M}\mathbf{J}_{12}\right).
\end{align}
In what follows, the freedom to add an integer multiple of $2\pi$ times a generator of rotation will be used in order to satisfy the requirement that for anti-de Sitter spacetime ($M=-1\,, J=0$) there should be no curvature singularity at the origin. This criterion implies adopting, instead of \eqref{WMAdS0},  the equivalent expression 
\begin{align} \label{WMAdS}
    \mathcal{W}_{M<0,J=0}=\exp\left(-2\pi (1- \sqrt{-M})\mathbf{J}_{12}\right),
\end{align}
since the generator $\mathbf{J}_{12}$ corresponds to a rotation in the $(1-2)$ plane. From \eqref{WMAdS} the torsion is found to be zero and the only nonvanishing curvature component for the static case is given by
\begin{align} \label{negative_sourceJ=0}
    R^{12}+\frac{e^{1}e^{2}}{\ell^2}=-2\pi (1-\sqrt{-M}) \delta(\Sigma),
\end{align}
which ensures that for AdS$_3$ spacetime the source is absent.

It is important to note that the Wilson loops shrinking around $r=0$ in the cases with $M=-n^2$ and $J=0$, with $n\geq 2 \in \mathbb{N}$, are also equal to the identity. This does not mean that those cases are source-free, but instead they represent $n$-fold rotations around the singularity, as discussed following \eqref{W=expA}.

The occurrence of a trivial Wilson loop, in the limit $r\to 0$, for static cones with $M=-n^2$ can be generalized also for the spinning ones. In fact, if the conserved charges are chosen as 
\begin{align} \label{MJ-BPS}
    M=-(m_+^2+m_-^2),\: J= 2  \ell \, m_+ m_-,
\end{align}
where $m_{\pm} \in \mathbb{Z}$, the Wilson loop \eqref{WM<0} reduces to the identity. The next section includes a discussion explaining the origin of these special values of $M$ and $J$. 

In the massless cases with a \textit{nonvanishing} angular momentum, the dreibein \eqref{frameM0} and Lorentz connection components \eqref{frame-sc-M0w01} and \eqref{frame-sc-M0w12} are considered yielding the following Wilson loop at $r=0$
\begin{align}
    \mathcal{W}_{M=0, J\neq 0}=\exp\left(-2\pi\left[\frac{1}{4}(\sgn(J) \mathbf{J}_{03}+\mathbf{J}_{23})-\frac{J}{\ell}\mathbf{J}_{01}-\frac{|J|}{\ell}\mathbf{J}_{12}\right]\right).
\end{align}
Then, the relevant curvature and torsion components are
\begin{align}
    R^{01}+\frac{e^{0}e^{1}}{\ell^2}=&-2\pi \frac{J}{\ell} \delta(\Sigma),\\
    R^{12}+\frac{e^{1}e^{2}}{\ell^2}=&-2\pi \frac{|J|}{\ell} \delta(\Sigma),\\
    T^{0}=&2\pi \frac{\ell \sgn(J)}{4}\delta(\Sigma)\\
    T^{2}=&2\pi \frac{\ell}{4}\delta(\Sigma).
\end{align}
Finally, the case $M=J=0$ is handled with the frame given by Eqs. (\ref{eMJ=0}) and  (\ref{wMJ=0}). The Wilson loop at a finite $r$ is then 
\begin{align}
    \mathcal{W}_{M=0,J=0}=\exp \left( -2 \pi \frac{r}{\ell} ( \mathbf{J}_{23} - \mathbf{J}_{12})  \right),
\end{align}
which, in the limit $r\to 0$ reduces to the identity, although this does not imply the absence of a curvature singularity. The nature of this singularity is different from that of the conical case, as will be discussed in the following section. 

\subsection{Wilson loop and the identification Killing vector} 

An interesting relation among the Wilson loop enclosing the singularity and the identification Killing vector $\mathbf{K}$ displayed in Table \ref{table:tablevectors} can be established from \eqref{wilsontheta}. It is possible to verify that for a circle of radius $r$, the Wilson loop corresponding to the geometry generated by $\mathbf{K}$, dubbed as  $\mathcal{W}_{\mathbf{K}}(r)$, takes the form
\begin{align}\label{generalWilson}
    \mathcal{W}_{\mathbf{K}}(r)=\mathcal{V}_{\mathbf{K}}(r)\,e^{-2 \pi \mathbf{K}}\,\mathcal{V}_{\mathbf{K}}^{-1}(r), 
\end{align}
where $\mathcal{V}_{\mathbf{K}}(r)$ is an element of $SO(2,2)$, such that $\mathcal{W}_{\mathbf{K}}$ is finite at $r=0$. The above equation explicitly shows the role of each identification Killing vector in its corresponding Wilson loop.

\begin{figure}
    \centering
    \includegraphics[width=1.0\linewidth]{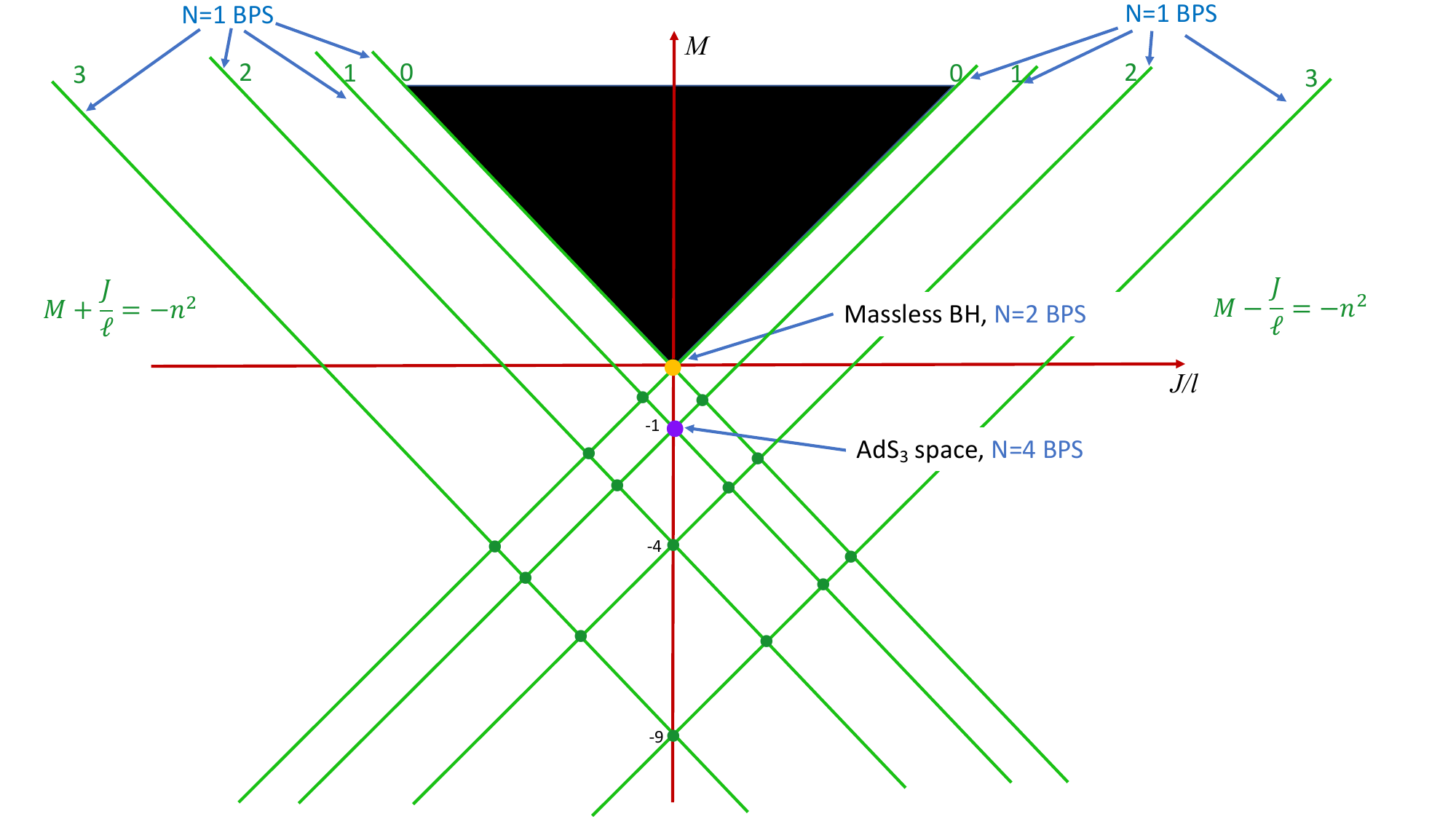}
    \caption{BPS configurations: Lines with $M\pm J/\ell = - n^2$ correspond to $N=1$ BPS states. The intersections of those lines are $N=2$ BPS, except for AdS$_3$ that is $N=4$ BPS.}
    \label{fig:fig2}
\end{figure}

Since the identification Killing vector for conical singularities is given by
\begin{align}\label{KGCS}
  \mathbf{K}= \lambda_+ \mathbf{J}_{21}+\lambda_- \mathbf{J}_{30},
\end{align}
namely, it is composed by two rotations, the property \eqref{generalWilson} proves to be particularly useful for a class of conical singularities where $\lambda_{\pm}=m_{\pm}\in \mathbb{Z}$. In this case the exponential $\exp(-2 \pi \mathbf{K})$ becomes the identity  and the Wilson loop \eqref{generalWilson} is trivial (for all $r$). This result is not obvious from the involved expression shown in \eqref{WM<0}, which holds just at $r=0$. In terms of the mass and momentum angular, the condition $\lambda_{\pm}=m_{\pm}\in \mathbb{Z}$ is translated as 
\begin{align}
    M=-(m_+^2+m_-^2),\: J= 2  \ell \, m_+ m_-.
\end{align}
 The conical naked singularities for these values of the conserved charges $M$ and $J$ have been identified as BPS configurations \cite{GMYZ}, which are shown in Fig. \ref{fig:fig2}. Note that for these BPS configurations the Wilson loop is trivial for any radius except for the case $(M,J)=(0,0)$. This result reinforces the idea that the massless BTZ black hole geometry is of a different nature from the other conical double BPS configurations. 

\section{Concluding remarks}\label{discusion} 

The nature of the central singularity of the BTZ geometries was analyzed using the holonomy operator corresponding to an infinitesimal loop around the central singularity. The study considered the Wilson loop for the Lorentz and the AdS$_3$ connection. The AdS$_3$ Wilson loop offers the advantage of obtaining information not only from the Lorentz curvature but also from the torsion, which are among the components of the AdS$_{3}$ curvature two-form. The analysis considered the complete BTZ family, that is, all values of mass $M$ and angular momentum $J$ in the metric \eqref{BTZ}.

In general, non-trivial Wilson loops are found in the zero-radius limit, which reveal non-trivial topologies as was mentioned in \cite{BHTZ}, or alternatively, the presence of delta-like singularities at the origin in the curvature and torsion two-forms. Remarkably, the Wilson loop reduces to the identity in the spinning cones cases where  $M\pm J/\ell=-n_{\pm}^2$, with $n_{\pm}\in \mathbb{N}$, which are exactly the BPS states mentioned in \cite{GMYZ}. Although these special geometries have a trivial Wilson loop, they are spinning  cones and posses a central singularity. In summary, except for the AdS$_{3}$ spacetime ($M=-1$, $J=0$), all BTZ geometries have a central singularity which is not revealed by local properties for $r\neq 0$. 

It is interesting to examine how the singularities arise from the identification process. In the embedding over the pseudosphere in $\mathbb{R}^{(2,2)}$, the surface with constant time and $r=0$ is a line. On the other hand, for the spacetime in 2+1 dimensions, the surface with constant time and $r=0$ is a point. Then, in principle, the mapping between the pseudosphere and the 2+1 dimensional spacetime at $r=0$ is not one-to-one in the generic case. This helps to understand the presence of a singularity\footnote{The AdS$_3$ is a special case in which the quotient space is smooth at $r=0$.} at $r=0$ and also the nontrivial holonomies found.

As explained in Section \ref{3.1}, the presence of rotation generators, whose exponential could lead to the identity, introduces an ambiguity in the equation \eqref{W=expA}, which relates the Wilson loop with the AdS$_{3}$ curvature.  This ambiguity is fixed by imposing \eqref{f=A}. Although \eqref{f=A} solves \eqref{W=expA}, it seems to be not the general answer, if any is possible to be found. Then, the expressions of the curvature and torsion displayed in Section \ref{3.2} could be generalized by means of a comprehensive treatment of equation \eqref{W=expA}. 

It was found that if $\lambda_+$ and $\lambda_-$ are integers, the identification Killing vector $\mathbf{K}= \lambda_+ \mathbf{J}_{21}+\lambda_- \mathbf{J}_{30}$ yields spinning cones, where the Wilson loops are trivial independent of the size of the loop.
Those configurations, recently discussed in \cite{GMYZ}, happen to be BPS states with angular excesses. The physical consequences emerging from the relation between these BPS states and the triviality of their associated Wilson loops were not addressed in this work, but they deserve attention.

The existence of a nonvanishing energy-momentum tensor poses the question about the nature of the source that produces this stress-energy tensor. For conical singularities --static or rotating-- it is natural to assume point particles as their sources. The identification is less obvious in the black hole case and for the overspinning geometry. In all cases, it could be expected that the action principle should include a contribution for the source, as shown in \eqref{source}. This source in turn, could also have its own dynamical properties, involving its own kinetic term. 
How would the inclusion of these source terms in the action be reflected in a quantum scenario is an open question. Probably, a treatment analogous to that for the back reaction of the geometry to the presence of a quantum field \cite{CFMZ0,CFMZ1,CFMZ2,CFMZ3,BaakeZanelli}, would have to be revised.

\vskip 3mm
\noindent{\bf{Acknowledgements.}} 
The authors thank Gaston Giribet, Hideki Maeda, Olivera Mi\v{s}kovi\'{c} and Nahuel Yazbek for useful discussions. This work has been partially funded by the ANID FONDECYT grants 1201208, 1220862 and 1241835.

\appendix


\end{document}